\documentclass[twocolumn,showpacs,preprintnumbers,prb,amsmath,amssymb]{revtex4}
\makeatletter
\def\@dotsep{4,5}
\makeatother

\usepackage{graphicx}
\usepackage{dcolumn}
\usepackage{bm}
\usepackage{subfigure} 

\begin{document}

\preprint{}

\title{Critical cavity in the stretched fluid studied using square-gradient density-functional model with triple-parabolic free energy}

\author{Masao Iwamatsu}
\email{iwamatsu@ph.ns.musashi-tech.ac.jp}
\affiliation{
Department of Physics, General Education Center,
Tokyo City University (Musashi Institute of Technology),
Setagaya-ku, Tokyo 158-8557, Japan
}%


\date{\today}

\begin{abstract}
The generic square-gradient density-functional model with triple-parabolic free energy is used to study the stability of a cavity introduced into the stretched liquid.  The various properties of the critical cavity, which is the largest stable cavity within the liquid, are compared with those of the critical bubble of the homogeneous bubble nucleation.  It is found that the size of the critical cavity is always smaller than that of the critical bubble, while the work of formation of the former is always higher than the latter in accordance with the conjectures made by Punnathanam and Corti [J. Chem. Phys. {\bf 119}, 10224 (2003)] deduced from the Lennard-Jones fluids.  Therefore their conjectures about the critical cavity size and the work of formation would be more general and valid even for other types of liquid such as metallic liquid or amorphous.  However, the scaling relations they found for the critical cavity in the Lennard-Jones fluid are marginally satisfied only near the spinodal.  
\end{abstract}

\maketitle

\section{Introduction}
\label{sec:level1}
The bubble plays a crucial role in many natural processes as well as in industrial practice or even in some industrial hazards~\cite{Debenedetti96,Oxtoby92,Vinogradov08}. Therefore, one of the basic mechanisms of the formation of the bubble called homogeneous bubble nucleation (cavitations) has attracted continuous attention for many years from a fundamental point of view as well as from technological interests~\cite{Shen01,Muller02,Delale03,Neimark05}.  The {\it homogeneous nucleation} occurs within a bulk fluid that is under-saturated and is held at a pressure lower than its coexisting vapor pressure at the given temperature.  Then the fluid is said to be stretched.  Another mechanism of the formation is the {\it heterogeneous nucleation}, where the bubble formation is induced by impurities in the liquid or by the wall of the container of the liquid. 

A tiny embryo of bubble called the nucleus forms within the stretched fluid from thermal fluctuation.  The work of formation $W$ of such a tiny bubble is positive, and has a maximum $W^{*}$ as the function of the size or radius of the embryo.  The nucleus with this critical size corresponding to the maximum $W^{*}$ is called the critical nucleus.  Therefore, a bubble that is smaller than this critical size is unstable and will shrink and disappear immediately. As soon as the size of the embryo becomes larger than this critical size, it starts to increase its size as the larger bubble becomes more favorable energetically. Thus the formation of bubbles is the activation process. The central quantities of the nucleation phenomena is the nucleation rate $J$, which is the number of critical nuclei formed per unit time per unit volume.  Usually it is written in Arrhenius form
\begin{equation}
J = A \exp\left(-\frac{W^{*}}{k_{\rm B}T}\right),
\label{eq:1}
\end{equation}
where $A$ is a kinetic pre-exponential factor which is believed to be weakly dependent on temperature $T$, and $k_{\rm B}$ is the Boltzmann's constant. Therefore, the temperature dependence of the nucleation rate is controlled by the work of formation $W^{*}$.  If any impurities or walls exist this will assist in lowering the necessary work $W^{*}$, the nucleation occurs predominantly near the impurity or wall.  Then the nucleation is called heterogeneous nucleation. Otherwise it occurs uniformly within the bulk and is called homogeneous nucleation. 

Recently, Corti~\cite{Punnathanam02,Punnathanam03,Punnathanam04,Uline07} and coworkers have studied not only the bubble nucleation but the cavity or void formation within the stretched liquid using the Monte Carlo simulation~\cite{Punnathanam02}, the density functional method~\cite{Punnathanam03,Uline07} and the thermodynamic perturbation method~\cite{Punnathanam04} for the Lennard-Jones fluid.  They conjectured that
\begin{itemize}
\item {\it Conjecture 1}:The size of the critical cavity is a lower bound to the size of critical bubble.
\item {\it Conjecture 2}:The work of formation of the critical cavity is the upper bound to the work of formation of the critical bubble.
\end{itemize}
where the critical cavity is the largest cavity that can be created within the liquid without inducing vaporization~\cite{Punnathanam02}.  A larger cavity will immediately drive the system to phase separation into the vapor phase.  Therefore the critical cavity represents the stability limit of the metastable liquid that contains the largest cavity. Based on these observation, Corti and coworkers~\cite{Punnathanam03} have conjectured that the homogeneous bubble nucleation is induced by the homogeneous cavity nucleation.  The cavity nucleation could play an important role to induce subsequent bubble nucleation.  Then the {\it homogeneous} bubble nucleation could, in fact, be the {\it heterogeneous} nucleation where the {\it homogenously} nucleated cavity would play the role of a spherical impurity or surface to induce {\it heterogeneous} nucleation of bubble~\cite{Padilla01,Brykov06,Lum99,Qian07}. These conjectures, however, are based on the numerical results of the cavity in the Lennard-Jones system only.  Therefore, further confirmation of their findings using a more generic model is necessary.

In this paper, we will use a simple square-gradient density functional theory with a triple-parabolic free energy~\cite{Granasy00} to study the cavity introduced into the stretched liquid~\cite{Punnathanam02,Punnathanam03,Uline07}.  The density functional theory (DFT) is known to be more reliable than the classical nucleation theory (CNT)~\cite{Debenedetti96,Oxtoby92} in particular near the spinodal~\cite{Cahn59,Klein83,Unger84,Binder84,Wilemski04}.  The square gradient approximation to DFT is also known to be qualitatively correct not only for the liquid with short-ranged interatomic potential~\cite{Davis96,Li03} but also for the liquid with the long-ranged potential~\cite{McCoy79,Barrett06}. Further approximation using parabolic free energy make this square-gradient density functional theory more attractive and generic as it does not depend on the detailed form of interatomic potentials.  It has elucidated the various aspects of the critical bubble in stretched liquid~\cite{Iwamatsu93,Barrett97,Iwamatsu08}. For example, we~\cite{Iwamatsu08} studied a scaling rule found for the Lennard-Jones system by Shen and Debenedetti~\cite{Shen01} and found that the scaling rule is marginally satisfied. Furthermore, this generic model can also be used to model the complex liquid such as polymer~\cite{Muller02} or even liquid metal~\cite{Granasy00}.    

This paper is organized as follows. In Section II we present a short review of the triple-parabola model~\cite{Granasy00} to summarize the necessary formula.  In Section III, we will present the numerical results and discuss the implications of the results in light of the conjectures of Punnathanam and Corti~\cite{ Punnathanam03}.  Finally Section IV is devoted to the conclusion.

\section{Square-gradient density functional model with triple-parabolic free energy}
\subsection{Triple-Parabolic Free Energy}
Since the model has been explained in detail elsewhere~\cite{Iwamatsu08}, we will briefly review the model.  In the square-gradient density-functional model of the fluid, the free energy of the inhomogeneous fluid, such as the critical bubble in the stretched liquid is written as
\begin{equation}
W=\int \left(\Delta\omega(\phi)+c\left(\nabla \phi\right)^{2}\right) d^{3}{\bf r},
\label{eq:2}
\end{equation}
where $\phi=\rho/\rho_{l}$ is the order parameter that represents the density $\rho$ divided by the liquid density $\rho_{l}$.  In the triple-parabola model of Gr\'an\'asy and Oxtoby~\cite{Granasy00}, the local part of the free energy $\Delta\omega$ is given by
\begin{equation}
\Delta \omega(\phi) = \left\{
\begin{array}{ll}
\frac{\lambda_{0}}{2}\left(\phi-\phi_{0}\right)^{2}+\Delta\mu , \;\; & \phi<\phi_{A} \\
\frac{\lambda_{1}}{2}\left(\phi-\phi_{1}\right)^{2}-\Delta\mu\frac{\phi_{1}-\phi_{0}}{\phi_{2}-\phi_{0}}+\Delta\mu+\omega_{0}, & \\
& \phi_{A}\leq\phi \leq\phi_{B} \\
\frac{\lambda_{2}}{2}\left(\phi-\phi_{2}\right)^{2}, \;\; & \phi_{B}<\phi
\end{array}
\right.
\label{eq:3}
\end{equation}
which consists of three parabola with $\lambda_{0}, \lambda_{2} > 0$ and $\lambda_{1} < 0$ centered at the vapor density $\phi_{0}$, and at the free energy barrier $\phi_{1}$ (spinodal), and at the liquid density $\phi_{2}$, which we call "vapor", "spinodal" and "liquid" part of the free energy. Equation (\ref{eq:2}) correctly represents the necessary work of forming bubble or cavity from the uniform fluid with relative density $\phi_{2}$ since the free energy of uniform liquid is given by $\omega\left(\phi_{2}\right)=0$.

The curvature of parabola $\lambda_{0}$ and $\lambda_{2}$ are related to the compressibility of vapor and liquid phases, and $\Delta\mu$ is the  free energy difference between the liquid and the vapor.  Although $\Delta\mu$ represents in fact the pressure difference $\Delta P$ as Eq.~(\ref{eq:2}) is the grand potential of the open system~\cite{Oxtoby92}, we call $\Delta\mu$ chemical potential to make the comparison to the previous work~\cite{Shen01,Punnathanam02,Punnathanam03} easier since $\Delta P$ is proportional to $\Delta\mu$ through $\Delta P=\rho_{\rm l}\Delta\mu$~\cite{Oxtoby92,Shen01}.  We use the terminology "over-saturation" when $\Delta\mu$ is positive and "under-saturation" when $\Delta\mu$ is negative.  The stretched liquid in this study corresponds to the under-saturated liquid with negative $\Delta\mu<0$.

From the continuity of the free energy $\Delta \omega(\phi)$ at the boundary, the matching densities $\phi_{A}$ and $\phi_{B}$ are given by
\begin{eqnarray}
\phi_{A}&=&\frac{\lambda_{0}\phi_{0}+\lvert\lambda_{1}\rvert\phi_{1}}{\lambda_{0}+\lvert\lambda_{1}\rvert}, \nonumber \\
\phi_{B}&=&\frac{\lambda_{2}\phi_{2}+\lvert\lambda_{1}\rvert\phi_{1}}{\lambda_{2}+\lvert\lambda_{1}\rvert},
\label{eq:4}
\end{eqnarray}
while the location $\phi_{1}$ of the free energy barrier is given by
\begin{equation}
\phi_{1}=\frac{(p\phi_{0}-q\phi_{2})+\sqrt{pq\left(\phi_{0}-\phi_{2}\right)^{2}-2\Delta\mu(p-q)}}{p-q},
\label{eq:5}
\end{equation}
with
\begin{eqnarray}
p &=& \frac{\lambda_{0}|\lambda_{1}|}{\lambda_{0}+|\lambda_{1}|}, \nonumber \\ 
q &=& \frac{\lambda_{2}|\lambda_{1}|}{\lambda_{2}+|\lambda_{1}|}.
\label{eq:6}
\end{eqnarray} 

The liquid spinodal, where the metastable stretched liquid becomes unstable, occurs when the under-saturation is given by
\begin{equation}
\Delta\mu_{\rm spin}=-\frac{p}{2}\left(\phi_{2}-\phi_{0}\right)^{2}.
\label{eq:7}
\end{equation}
In contrast to the previous models~\cite{Shen01,Unger84,Binder84,Wilemski04} where the compressibility diverges continuously as the spinodal is approached ($\Delta\mu\rightarrow \Delta\mu_{\rm spin}$),  the compressibility remains finite until the spinodal point is reached in our triple-parabolic model as the curvature $\lambda_{0}$ and $\lambda_{2}$ is fixed.  In the usual $\phi^{4}$ model, the diverging~\cite{Klein83,Unger84,Kalikmanov04,Wood02} compressibility of metastable liquid phase
\begin{equation}
\lambda_{2} \propto \left(1 -\frac{\Delta\mu}{\Delta \mu_{\rm spin}}\right)^{1/2}
\label{eq:8}
\end{equation}
is predicted.  Therefore, the results obtained from this triple-parabolic free energy model are somehow artificial near the spinodal.

\begin{figure}[htbp]
\begin{center}
\includegraphics[width=0.85\linewidth]{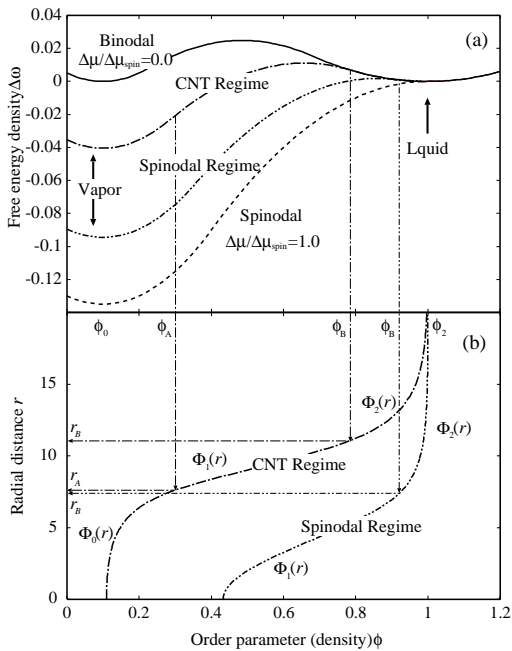}
\end{center}
\caption{
(a) The triple-parabolic free energy from CNT regime near the coexistence to the spinodal regime near the liquid spinodal for the case-I of section III (Table~\ref{tab:1}). (b) The corresponding critical bubble of homogeneous nucleation at the CNT regime ($\Delta\mu/\Delta\mu_{\rm spin}=0.3$) and at the spinodal regime ($0.7$).
}
\label{fig:1}
\end{figure}

In Fig.~\ref{fig:1}(a) we show typical shape of the triple-parabolic free energy $\Delta \omega$ together with the corresponding density profile of the critical bubble in Fig.~\ref{fig:1}(b).  The free energy $\Delta \omega$ shows a typical double-well form with the lower well represents the stable vapor phase and the upper well represents the metastable liquid phase.  The radii $r_{A}$ and $r_{B}$ are the matching radius that satisfies $\phi\left(r_{A}\right)=\phi_{A}$ and $\phi\left(r_{B}\right)=\phi_{B}$. Since the free energy consists of three parabolas corresponding to the vapor, spinodal and liquid parts, the density profile of the critical bubble also consists of three parts when $\Delta\mu/\Delta\mu_{\rm spin}=0.3$ near the coexistence.  However, as the under-saturation increases ($\lvert\Delta\mu\rvert$ becomes large) and it approaches the liquid spinodal $\Delta\mu_{\rm spin}$, the density profile consists of only the two parts corresponding to the spinodal and the liquid parts as the density never decreases down to the vapor density.  We use the terminology "CNT regime" for the former regime where the classical nucleation theory (CNT) is expected to be qualitatively correct, and "spinodal regime" for the latter where the spinodal nucleation~\cite{Debenedetti96,Unger84,Binder84,Wilemski04} is expected to occur.

\subsection{Density Profile of Critical Cavity and Bubble}

\subsubsection{CNT regime}
Density profile of spherically symmetric critical cavity and bubble can be obtained from the Euler-Lagrange (EL) equation $\delta W/\delta \phi=0$, which corresponds to the saddle point of the free energy surface.  This EL equation reads to the differential equation 
\begin{equation}
\frac{d^{2}\Phi_{i}}{dr^{2}}+\frac{2}{r}\Phi_{i}\pm \Gamma_{i}^{2}\Phi_{i}=0,\;\;\;i=0,1,2
\label{eq:9}
\end{equation}
for the three parabolas in Eq.~(\ref{eq:1}), where $\Gamma_{i}=\sqrt{\lvert\lambda_{i}\rvert/2c}$ and $\Phi_{i}(r)=\phi(r)-\phi_{i}$, and $+$ sign is used for $i=1$ and $-$ is used for $i=0, 2$ for $\pm$. These differential equations should be solved with boundary conditions:
\begin{eqnarray}
\Phi_{0}\left(r=r_{A}\right)&=&\Phi_{0A}=\phi_{A}-\phi_{0}, \nonumber \\
\Phi_{1}\left(r=r_{A}\right)&=&\Phi_{1A}=\phi_{A}-\phi_{1}, \nonumber \\
\Phi_{1}\left(r=r_{B}\right)&=&\Phi_{1B}=\phi_{B}-\phi_{1}, \label{eq:10} \\
\Phi_{2}\left(r=r_{B}\right)&=&\Phi_{2B}=\phi_{B}-\phi_{2} ,\nonumber \\
\Phi_{2}\left(r\rightarrow \infty \right) &=& 0 , \nonumber
\end{eqnarray}
together with the condition for the vapor density $\Phi_{0}$ given by
\begin{equation}
\Phi_{0}\left(r\rightarrow R^{+}\right)=-\phi_{0}\;\;\;\mbox{(Critical cavity)}
\label{eq:11}
\end{equation}
for the critical cavity with the radius $R$, and
\begin{equation}
\left.\frac{d\Phi_{0}}{dr}\right|_{r\rightarrow 0}=0\;\;\;\mbox{(Critical bubble)}
\label{eq:12}
\end{equation}
for the critical bubble. The boundary condition for the cavity in Eq.~(\ref{eq:11}) implies that the work of formation $W(R)$ of the cavity reaches a maximum value as the function of the cavity radius $R$ since~\cite{Reiss59}
\begin{equation}
\frac{\partial W}{\partial R}=4\pi R^{2}\phi\left(R^{+}\right)k_{\rm B}T=0
\label{eq:12x}
\end{equation}
and $\phi\left(R~{+}\right)=\Phi_{0}\left(R^{+}\right)+\phi_{0}=0$. This condition can also be interpreted as the onset of the drying transition of fluid surrounding the cavity~\cite{Lum99,Punnathanam02}.  The critical cavity is not only defined as the saddle point solution of Eq.(\ref{eq:9}) with the largest cavity radius $R$ but corresponds in fact to the stability limit as $\partial W/\partial R\geq 0$.  Therefore the critical cavity is the limit of the stability of the metastable liquid that contains the largest cavity.  Once the size of the cavity exceeds the size of the critical cavity, the stretched liquid immediately phase separates into the vapor phase.  On the other hand, the critical bubble corresponds to the intermediate highest energy saddle point of the reaction of the growth of the nucleus.  Once the size of the bubble exceeds that of the critical bubble, it increases indefinitely and will eventually lead to the phase transition of the whole system from the metastable liquid to the stable vapor phase.

In this CNT regime near the two-phase coexistence, both the matching radius $r_{A}$ and $r_{B}$ exist. When the under saturation $\Delta \mu (<0)$ is further increased, the matching radius $r_{A}$ approaches zero and disappears for the bubble solution (Fig.~\ref{fig:1}).  Then only the matching radius $r_{B}$ can exist.  Such a transition never happened to the critical cavity as the density always becomes the vapor density from the boundary condition Eq.~(\ref{eq:11}).  In the critical cavity, not only the matching radius $r_{B}$ but the radius $r_{A}$ exist up to the spinodal point.

The solutions of this EL equation Eq.~(\ref{eq:9}) for the critical cavity in the liquid ($i=2$) and the spinodal ($i=1$) part of the free energy are given by
\begin{eqnarray}
\Phi_{1}(r) &=& 
\csc\left(\Gamma_{1}\left(r_{A}-r_{B}\right)\right) \nonumber \\
&& \times (
-\Phi_{1B}r_{B}\sin\left(\Gamma_{1}\left(r-r_{A}\right)\right) \label{eq:13} \\
&&
+\Phi_{1A}r_{A}\sin\left(\Gamma_{1}\left(r-r_{B}\right)\right)
)/r ,  \nonumber \\
\Phi_{2}(r) &=& \Phi_{2B}r_{B}\exp\left(-\Gamma_{2}r+\Gamma_{2}r_{B}\right)/r , \nonumber
\end{eqnarray}
while the one in the vapor ($i=0$) part is given by
\begin{eqnarray}
\Phi_{0}(r) &=& -{\rm csch}\left(\Gamma_{0}\left(R-r_{A}\right)\right) \nonumber \\
&&\times\left(\Phi_{0A}r_{A} {\rm sinh}\left(\Gamma_{0}\left(r-R\right)\right)\right. 
\label{eq:14} \\
&& \left.+\phi_{0}R{\rm sinh}\left(\Gamma_{0}\left(r-r_{A}\right)\right)\right)/r  \nonumber
\end{eqnarray}
The expression for the critical bubble can be obtained by setting $R=0$. 

Finally, the matching radii $r_{A}$ and $r_{B}$ are determined from the simultaneous equation
\begin{eqnarray}
\left.\frac{d\Phi_{2}}{dr}\right|_{r=r_{B}} &=& \left.\frac{d\Phi_{1}}{dr}\right|_{r=r_{B}}, \nonumber \\
\left.\frac{d\Phi_{1}}{dr}\right|_{r=r_{A}} &=& \left.\frac{d\Phi_{0}}{dr}\right|_{r=r_{A}},
\label{eq:15} 
\end{eqnarray}
where $\Phi_{1}$ are the functions of both $r_{A}$ and $r_{B}$.  This simultaneous equation can be solved numerically using standard algorithms such as the Newton-Raphson method for both the critical cavity and the bubble.

\subsubsection{Spinodal regime}

In this regime near the spinodal, the density of vapor bubble remains high even at the center of the bubble.  Therefore the bubble remains within the liquid well of Fig.~\ref{fig:1}(a), and  $r_{A}$ for the critical bubble becomes zero (see Fig.~\ref{fig:1}).   Therefore the $i=0$ part of the free energy density in Eq.~(\ref{eq:3}) and its solution $\Phi_{0}$ disappears.  Then the boundary condition Eqs.~(\ref{eq:11}) and (\ref{eq:12}) for $\Phi_{0}$ should be replaced by the same condition for $\Phi_{1}$.  In fact, such a situation can happen only for the homogeneous bubble nucleation~\cite{Iwamatsu08}, but it has never happened to the critical cavity as the density always goes down to zero at the cavity boundary from Eq.~(\ref{eq:11}) and the vapor solution Eq.~(\ref{eq:14}) is always necessary to represents cavity density.  However, for the sake of completeness, we will present the cavity solution in the spinodal regime as well.

The solution for the EL equation for $\Phi_{2}$ is the same as Eq.~(\ref{eq:12}), but the one for $\Phi_{1}$ now read
\begin{eqnarray}
\Phi_{1}(r) &=& -{\csc}\left(\Gamma_{1}\left(R-r_{B}\right)\right) \nonumber \\
&& \times \left(\Phi_{1B}r_{B}\sin\left(\Gamma_{1}\left(r-R\right)\right) \right. \label{eq:16} \\
&& \left.+\phi_{1}R\sin\left(\Gamma_{1}\left(r-r_{B}\right)\right)\right)\nonumber
\end{eqnarray}
for the critical cavity in the spinodal regime.  Again the expression for the critical bubble is recovered by setting $R=0$.

In this case, the matching radius $r_{B}$ is simply determined from the equation
\begin{equation}
\left.\frac{d\Phi_{2}}{dr}\right|_{r=r_{B}} = \left.\frac{d\Phi_{1}}{dr}\right|_{r=r_{B}}. 
\label{eq:17}
\end{equation}
In particular, Eq.~(\ref{eq:17}) for the critical bubble does not depend on the under-saturation $\Delta\mu$.  Therefore, the matching radius $r_{B}$ is constant in the spinodal regime for the critical bubble~\cite{Iwamatsu08}. 

In Fig.~\ref{fig:1}(b), we showed the typical density profiles of critical bubble in the CNT regime and in the spinodal regime.  The critical bubble is larger in the CNT regime than that in the spinodal regime.  However, the density difference between the inside and the outside of the bubble becomes smaller in the spinodal regime than that in the CNT regime.  Since the length scale $1/\Gamma_{i}=\sqrt{2c/|\lambda_{i}|}$ does not depend on the under-saturation $\Delta\mu$, the shape of the critical bubble, in particular, decay length of the tail, or the interfacial width look almost constant.

\subsection{Work of Formation of Critical Cavity and Bubble}
\subsubsection{CNT regime}

Once we know the density profile of bubble and cavity, it is straightforward to calculate the work of formation $W^{*}$ of the critical cavity and bubble.  Since, the results for the critical bubble have already been presented elsewhere~\cite{Iwamatsu08}, we only show the results for the cavity.  To this end, we can use the formula
\begin{equation}
W^{*} =\int_{R}^{\infty}4\pi r^{2}\left(\Delta\omega-\frac{1}{2}\phi\frac{\partial \Delta\omega}{\partial \phi}\right)dr
\label{eq:18}
\end{equation}
derived by Cahn and Hilliard~\cite{Cahn59}.  Using this transformation, we can evade the singularity of $\nabla \phi$ in Eq.~(\ref{eq:2}).  Like previous authors~\cite{Punnathanam02,Punnathanam03,Uline07}, we regard the critical cavity as the external hard wall and omit the contribution from $r<R$ in Eq.~(\ref{eq:18}).

Inserting the density profile Eq.~(\ref{eq:12}) in the CNT regime, we can calculate the integral in Eq.~(\ref{eq:18}) analytically, and we obtain
\begin{equation}
W=W_{0}+W_{1}+W_{2},
\label{eq:19}
\end{equation}
where
\begin{eqnarray}
W_{1}&=& \frac{4\pi}{3}\left(r_{B}^{3}-r_{A}^{3}\right)\left(-\Delta\mu\frac{\phi_{1}-\phi_{0}}{\phi_{2}-\phi_{0}}+\Delta\mu+\omega_{0}\right)  \nonumber \\
&&+\frac{2\pi|\lambda_{1}|\phi_{1}}{\Gamma_{1}^{2}}\left(-\Phi_{1A}r_{A}+\Phi_{1B}r_{B}\right. \label{eq:20} \\
&&+\left.\Gamma_{1}\left(\Phi_{1A}r_{A}^{2}+\Phi_{1B}r_{B}^{2}\right)\cot\left(\Gamma_{1}\left(r_{A}-r_{B}\right)\right)\right. \nonumber \\
&&-\left.\Gamma_{1}\left(\Phi_{1A}+\Phi_{1B}\right)r_{A}r_{B}{\rm csc}\left(\Gamma_{1}\left(r_{A}-r_{B}\right)\right)\right), \nonumber \\
W_{2} &=& -\frac{2\pi\lambda_{2}\phi_{2}\Phi_{2B}r_{B}}{\Gamma_{2}^{2}}\left(1+r_{B}\Gamma_{2}\right), \nonumber 
\end{eqnarray}
for both the critical cavity and critical bubble, while
\begin{eqnarray}
W_{0}&=&\frac{4\pi}{3}r_{A}^{3}\Delta\mu+\frac{2\pi\lambda_{0}\phi_{0}}{\Gamma_{0}^{2}}\left(\pi_{0}R+\Phi_{0A}r_{A}\right. \nonumber \\
&& +\Gamma_{0}\left(\phi_{0}R^{2}+\Phi_{0A}r_{A}^{2}\right){\rm coth}\left(\Gamma_{0}\left(R-r_{A}\right)\right) \label{eq:21} \\
&& \left.+\Gamma_{0}\left(\phi_{0}-\Phi_{0A}\right)r_{A}R{\rm csch}\left(\Gamma_{0}\left(R-r_{A}\right)\right)\right) \nonumber
\end{eqnarray}
for the critical cavity. The expression of $W_{0}$ for the critical bubble is recovered when $R=0$.  Note that $\Delta\mu<0$ for the critical bubble in the under-saturated liquid.

\subsubsection{Spinodal regime}

Since, we only have the solution $\Phi_{2}$ in Eq.~(\ref{eq:13}) and $\Phi_{1}$ given by Eq.~(\ref{eq:16}) for the critical bubble in the spinodal regime, we have 
\begin{equation}
W=W_{1}+W_{2},
\label{eq:22}
\end{equation}
where $W_{2}$ is given by Eq.~(\ref{eq:20}) but $W_{1}$ is given by
\begin{eqnarray}
W_{1} &=& \frac{4\pi}{3}r_{B}^{3}\left(-\Delta\mu\frac{\phi_{1}-\phi_{0}}{\phi_{2}-\phi_{0}}+\Delta\mu+\omega_{0}\right) \nonumber \\
&&+\frac{2\pi|\lambda_{1}|\phi_{1}}{\Gamma_{1}^{2}}\left(\phi_{1}R+\Phi_{1B}r_{B} \right.
\label{eq:23} \\
&& +\Gamma_{1}\left(-\phi_{1}R^{2}+\Phi_{1B}r_{B}^{2}\right)\cot\left(\Gamma_{1}\left(R-r_{B}\right)\right)  \nonumber \\
&& \left. +\Gamma_{1}\left(\phi_{1}-\Phi_{1B}\right)r_{B}R{\rm csc}\left(\Gamma_{1}\left(R-r_{B}\right)\right)\right) \nonumber
\end{eqnarray}
for the critical cavity. The expression of $W_{1}$ for the critical bubble is recovered when $R=0$.  However, this equation Eq.~(\ref{eq:22}) will not be used for the critical cavity as it corresponds to the unphysical solution with higher free energy.

\section{Numerical Results and discussions}

In order to check the validity of the two conjectures by Punnathanam and Corti~\cite{Punnathanam03} mentioned in the introduction derived from the numerical results from the Lennard-Jones fluid, we use this generic square-gradient density-functional model with triple-parabolic free energy and compare the various properties of the critical cavity and bubble.  We use several typical free energy parameters used before~\cite{Iwamatsu08} to confirm the results further.  The three sets of the free energy parameters used are summarized in Table~\ref{tab:1}.

\begin{table}[htbp]
\caption{
Three sets of free energy parameters used in this work to study the various properties of critical cavity and bubble.
}
\label{tab:1}
\begin{center}
\begin{tabular}{c|cccccc}
\hline
model & $c$ & $\phi_{0}$ & $\phi_{2}$ &  $\lambda_{0}$ & $\lambda_{1}$ & $\lambda_{2}$ \\
\hline
case-I & 1.0 & 0.1 & 1.0 & 1.0 & -0.5 & 0.3 \\
case-II & 1.0 & 0.1 & 1.0 & 1.0 & -1.0 & 0.7 \\
case-III & 1.0 & 0.1 & 1.0 & 1.0 & -2.0 & 0.9 \\
\hline
\end{tabular}
\end{center}
\end{table}

\begin{figure}[htbp]
\begin{center}
\includegraphics[width=0.85\linewidth]{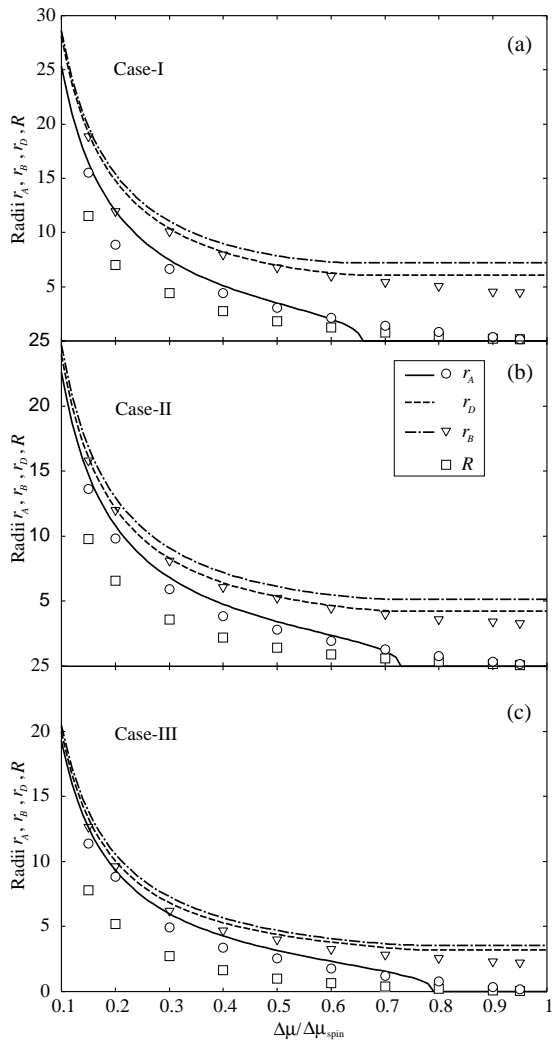}
\end{center}
\caption{
The matching radii $r_{A}$ (circle) and $r_{B}$ (triangle) as well as the critical radius $R$ (square) of the critical cavity.  Also shown are the matching radii $r_{A}$ (solid curve) and $r_{B}$ (chain curve) as well as the dividing radius $r_{\rm D}$ (dashed curve) of the critical bubble for the case-I to III.}
\label{fig:2}
\end{figure}

Figure \ref{fig:2} compares the matching radius $r_{A}$ and $r_{B}$ as well as the critical radius $R$ of the critical cavity and bubble~\cite{Iwamatsu08} as the functions of the scaled under-saturation $\Delta \mu/\Delta \mu_{\rm sp}$ for the case-I to III. The critical radius is the maximum radius $R$ for which the saddle point solution for Eq.~(\ref{eq:9}) can exist.  We manually increased the radius $R$ of the cavity and monitor the stability of the solution and determined the maximum radius $R$.

Figure \ref{fig:2} also shows the matching radius $r_{A}$ and $r_{B}$ of the critical bubble reported before~\cite{Iwamatsu08} as the functions of the scaled under-saturation $\Delta \mu/\Delta \mu_{\rm sp}$ together with the radius $R$ of the critical cavity for the case-I to III. The matching radius $r_{A}$ becomes zero and the density profile is represented by Eq.~(\ref{eq:16}) as the under-saturation enters the spinodal regime, while the matching radius $r_{B}$ survives and it becomes constant and independent of the under-saturation in the spinodal regime as predicted from Eq.~(\ref{eq:16}).  We also show the equimolar dividing radius $r_{D}$ of the critical bubble determined from~\cite{Iwamatsu08}
\begin{equation}
\int_{0}^{r_{D}}\left(\phi(r)-\phi_{\rm org}\right)4\pi r^2 dr
=\int_{r_{D}}^{\infty}\left(\phi_{2}-\phi(r)\right)4\pi r^2 dr ,
\label{eq:24}
\end{equation}
where
\begin{equation}
\phi_{\rm org}=\phi(r\rightarrow 0)
\label{eq:25}
\end{equation}
is the density at the origin of the bubble.  The dividing radius $r_{D}$ is constant as the matching radius $r_{B}$ in the spinodal regime. In fact, these constant radii are due to the fact that the parameter $\lambda_{2}$ which represents the compressibility of the metastable liquid remains constant.  Therefore the constant radius in the spinodal regime up to the spinodal is merely the artifact of our model.  Rather, the diverging radius is expected as the compressibility of metastable liquid diverges at the spinodal~\cite{Debenedetti96,Unger84,Binder84} according to Eq.~(\ref{eq:8}).  In contrast, the critical radius $R$ of the critical cavity converges to zero at the spinodal.

In contrast to the critical bubble, the matching radius $r_{A}$ of the critical cavity does not disappear in the spinodal region but decrease monotonically and approaches zero at the spinodal. This is due to the boundary condition Eq.~(\ref{eq:11}) and the fact that the density always decreases down to a vacuum.  The density profile of critical cavity remains within the CNT regime down to the spinodal.  Since the matching radius $r_{A}$ approaches zero at the spinodal, the critical radius $R$ also approaches zero as $R$ should always be smaller than the matching radius $r_{A}$ of the critical cavity. The matching radius $r_{B}$ remains finite even at the spinodal, which indicates the finite width of the liquid-vapor interface of the cavity.  No matter what the definition of the size of the critical cavity and bubble, the size of critical cavity represented by $R$ or $r_{B}$ seems the lower bound of the corresponding radius such as the dividing radius $r_{D}$ and the matching radius $r_{B}$ of the critical bubble as shown in Fig.~\ref{fig:2}.

\begin{figure}[htbp]
\begin{center}
\includegraphics[width=0.85\linewidth]{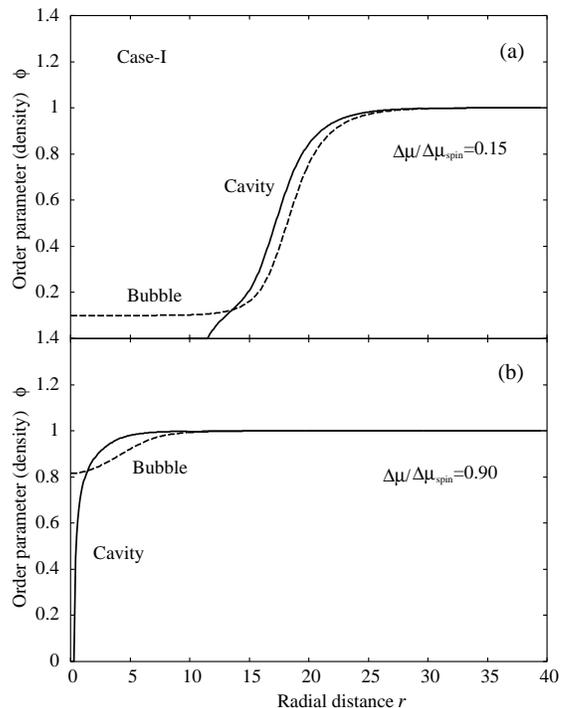}
\\
\end{center}
\caption{
(a) The density profile (order parameter) of the critical cavity compared with that of the critical bubble in the CNT regime of the low under-saturation $\Delta \mu/\Delta \mu_{\rm spin}=0.15$ for the Case-I.  (b) The density profiles in the spinodal regime of the high under-saturation $\Delta \mu/\Delta \mu_{\rm spin}=0.90$. 
}
\label{fig:3}
\end{figure}

In fact, Figure~\ref{fig:3} clearly indicates that not only the radius $R$ but the overall profile of the critical cavity is smaller than the critical bubble.  In the CNT regime of the low under-saturation, in particular, the density profile of critical cavity is similar to the corresponding critical bubble. Not only the size but also the width of the liquid-vapor interface is similar magnitude.   However, the liquid within the critical bubble is more pushed outward compared to the liquid within the critical cavity even though the density becomes zero for $r<R$ for the critical cavity (Fig.~\ref{fig:3}(a)).  The typical radius of the bubble $r_{D}$ or $r_{B}$ is always larger than $R$ or $r_{B}$ for the critical cavity as shown in Fig.~\ref{fig:2}.

The situation is less clear in the spinodal regime as the density profile of the critical cavity and bubble are totally different (Fig.~\ref{fig:3}(b)).  The density of cavity goes to zero at the critical radius $R$ while the density of the critical bubble within the bubble is almost the same as that of the liquid.  Therefore the width of the liquid-vapor interface is also different, and the interface of bubble is much wider than that of the cavity.  Yet, we can say that the cavity is smaller than the bubble as the typical size $r_{B}$ of the cavity, for example, is always smaller than that of the bubble.  In particular, not only the typical size $r_{A}$ but the critical radius $R$ for the critical cavity approaches zero at the spinodal, while the dividing radius $r_{D}$ for the critical bubble remains finite~\cite{Iwamatsu08}.

The density profile of critical cavity is the saddle point solution of the EL equation Eq.~(\ref{eq:9}) with the lowest free energy and maximum radius $R$.  As we increase the radius of cavity, the work of formation $W$ increases monotonically because $\partial W/\partial R\geq 0$ from Eq.~(\ref{eq:12x}).  As soon as the radius of the cavity exceeds the critical radius $R$ that corresponds to $\partial W/\partial R =0$, the free energy jumps to the higher energy "excited-state" of Eq.~(\ref{eq:9}) and the oscillating density profile appears.  The origin of this oscillation is not the same as the one observed in the previous Monte Carlo simulation~\cite{Punnathanam02} as the latter comes from the excluded volume effect.  Our square-gradient approximation cannot include such an excluded volume effect, and our oscillatory solutions are not real.  Rather such an oscillatory solution indicates that the stable solution which smoothly connects to the liquid density at infinity cannot exist~\cite{Punnathanam03}.  The cavity with a radius larger than the critical radius $R$ induces more density oscillation and a higher free energy.  Therefore, we interpreted the appearance of oscillatory solution as the limit of stability of the cavity.  The solution with maximum radius $R$ without density oscillation corresponds to the critical cavity which is the cavity of the maximum size that can be accommodated into the stretched fluid.  

\begin{figure}[htbp]
\begin{center}
\includegraphics[width=0.85\linewidth]{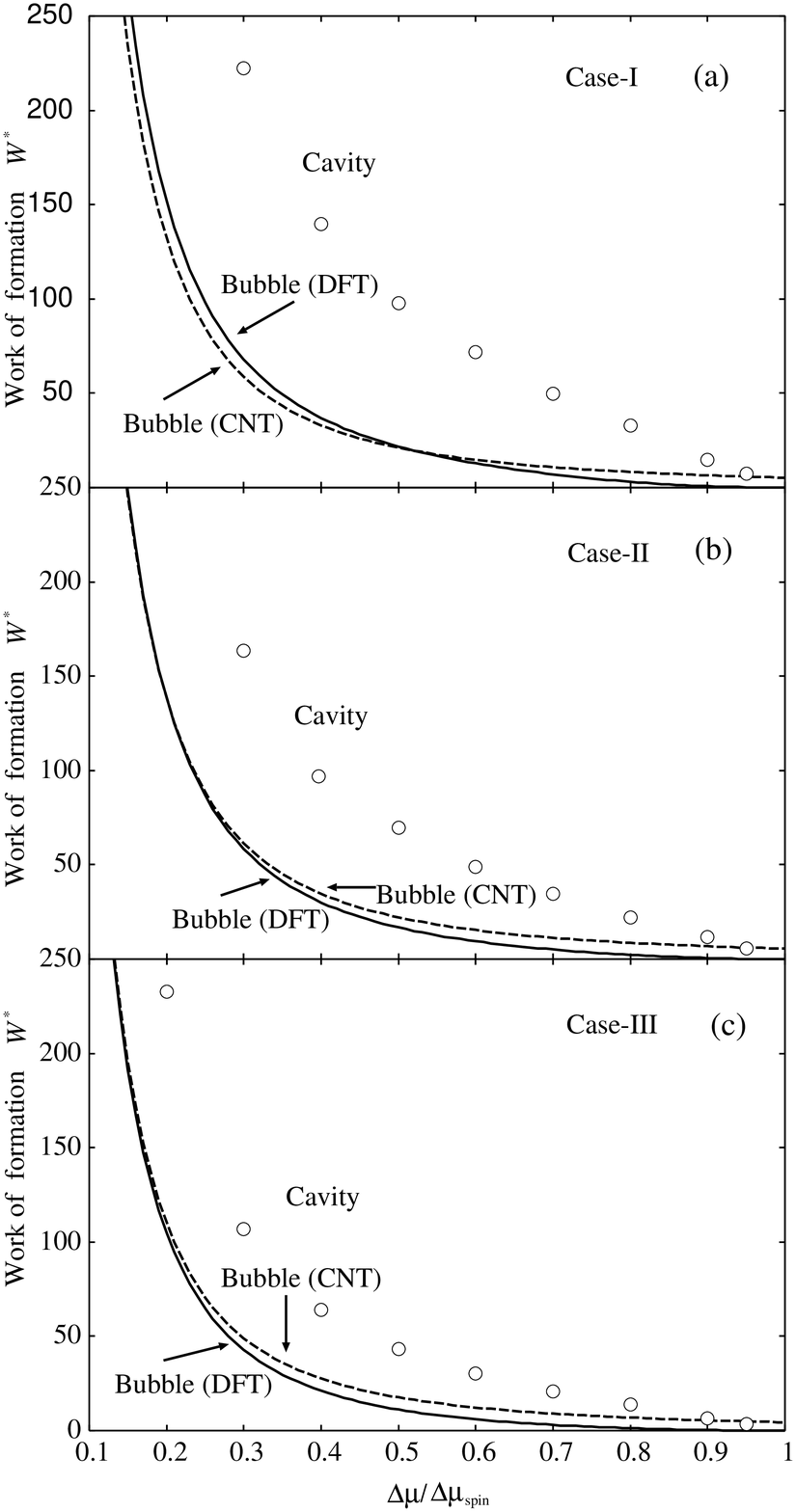}
\end{center}
\caption{
The work of formation $W^{*}$ of the critical cavity (circle) compared with that of the critical bubble calculated from DFT (solid line) as well as that from CNT (dashed line) for the case-I to III.
}
\label{fig:4}
\end{figure}

The work of formation $W^{*}$ of the critical bubble and cavity are compared in Fig.~\ref{fig:4}.  The work of formation for the cavity is always much larger than the critical bubble.  The work of formation of critical cavity approaches zero at the spinodal. In contrast, the work of formation of critical bubble calculated from the CNT erroneously remains finite~\cite{Li03,Kalikmanov04} even at the spinodal.  Although both the work of formation of cavity and bubble from the DFT approaches zero, the former seems always larger than the latter in accordance with the results from a more sophisticated non-local DFT for the Lennard-Jones fluid~\cite{Punnathanam03}. 

From the results for the case-I to III in Figs.~\ref{fig:2} and \ref{fig:4}, it seems fair to say that the size of the critical cavity is always smaller than the size of the critical bubble, while the work of formation of the critical cavity is always larger than that of the critical bubble in accordance with the {\it conjecture 1} and {\it conjecture 2} of Punnathanam and Corti~\cite{Punnathanam03} mentioned in the introduction.  In particular, the work of formation for the critical cavity is much higher than that of the critical bubble. 

\begin{figure}[htbp]
\begin{center}
\includegraphics[width=0.85\linewidth]{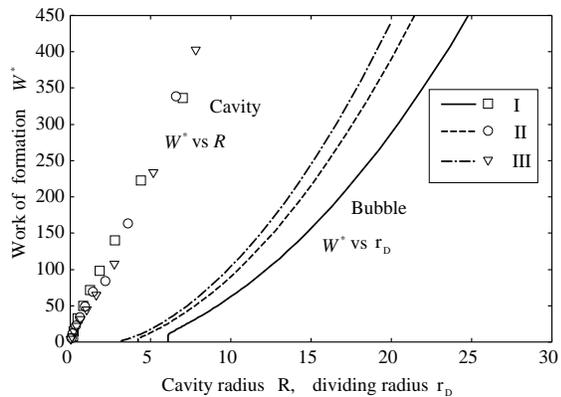}
\end{center}
\caption{
The work of formation $W^{*}$ of the critical cavity (circle) and bubble (solid line) as the function of the cavity radius $R$ or the dividing radius $r_{D}$ of the bubble.
}
\label{fig:5}
\end{figure}

Figure \ref{fig:5} shows the work of formation of the critical cavity and the bubble as the function of the cavity radius $R$ of the critical cavity or the dividing radius $r_{D}$ of the critical bubble. The work of formation $W^{*}$ for the critical cavity scales with the parameter $R$, while the one for the critical bubble does not scale with the dividing radius $r_{D}$.  

\begin{figure}[htbp]
\begin{center}
\subfigure[Scaled radius $R$ and the matching radius $r_{B}$ of critical cavity]{\includegraphics[width=0.90\linewidth]{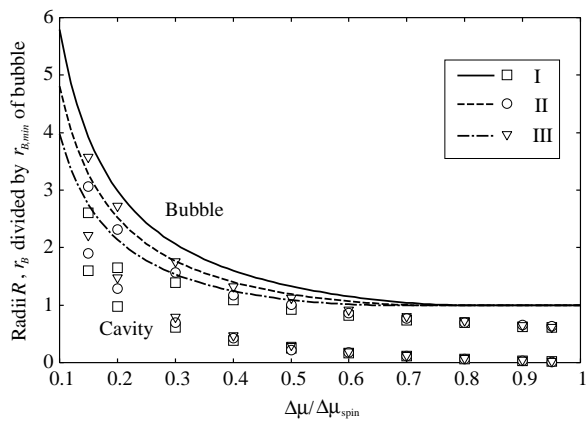}
\label{fig:6xa}}
\subfigure[Scaled work of formation $W^{*}$ of critical cavity]{\includegraphics[width=0.90\linewidth]{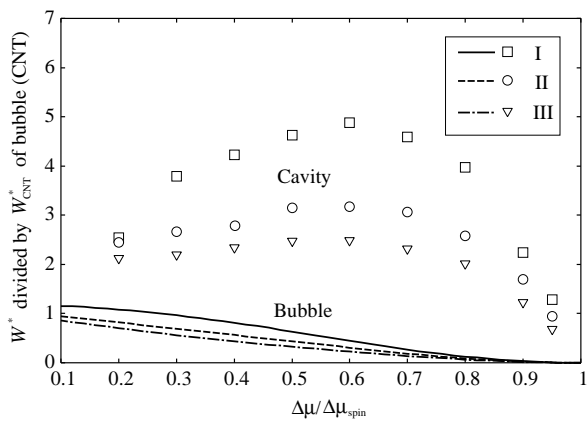}
\label{fig:6xb}}
\end{center}
\caption{
(a)Scaled radius $R$ and the dividing radius $r_{B}$ of critical cavity as the function of the scaled undersaturation $\Delta\mu/\Delta\mu_{\rm spin}$.  (b) Scaled work of formation $W^{*}$ of critical cavity as the function of the scaled undersaturation $\Delta\mu/\Delta\mu_{\rm spin}$.
}
\label{fig:6}
\end{figure}

In the previous paper~\cite{Iwamatsu08}, we have checked the scaling rule for the various quantities of the critical bubble of our square-gradient DFT-model as a function of scaled under saturation $\Delta\mu/\Delta\mu_{\rm spin}$.  In contrast to the Lennard-Jones fluid where such a scaling rule is indeed valid~\cite{Shen01}, we found that although some quantities show almost perfect scaling relations near the spinodal, the work of formation divided by the value deduced from the CNT shows no scaling~\cite{Iwamatsu08}.  Similar scaling rules for cavity are also found in the Lennard-Jones fluid by Punnathanam and Corti~\cite{Punnathanam03}.  In order to check the scaling rule of cavity in our DFT-model we have plotted the critical radius $R$ and the matching radius $r_{B}$ of the critical cavity divided by $r_{B}$ of the critical bubble as the function of scaled undersaturation $\Delta\mu/\Delta\mu_{\rm spin}$ in Fig.~\ref{fig:6}(a).  The critical radius $R$ as well as the matching radius $r_{B}$ of cavity show similar scaling relations in particular near the spinodal to the various radii of critical bubble~\cite{Iwamatsu08} of the same DFT-model.  However, Fig.~\ref{fig:6}(b) clearly indicates that the work of formation $W^{*}$ of the critical cavity as the function of the scaled undersaturation $\Delta\mu/\Delta\mu_{\rm spin}$ shows no scaling again~\cite{Iwamatsu08}.  

We should note that the cavity and bubble in Fig.~\ref{fig:5} represents critical cavity and bubble exactly at the saddle point as we consider the open system within the framework of the grand canonical ensemble.  Therefore this diagram is {\it not} for the work of formation $W(n,r)$ of cavity or bubble with arbitrary number of molecule $n$ with arbitrary size $r$ or  $R$ in the canonical ensemble~\cite{Punnathanam03,Uline07}.  Although this diagram is for the critical cavity or bubble, it is expected that the work of formation of any cavity or bubble with arbitrary size in the canonical ensemble would show a similar trend as the function of its radius $R$.  Then, it is expected that the (sub-critical) cavity, which is smaller than the critical cavity and is not large enough to induce instability, would always exist near the spinodal as both the work of formation $W^{*}$ and the size of the critical cavity $R$ of even the largest critical cavity approaches zero at the spinodal. 

In our DFT-model, the size of the critical bubble approaches finite minimum size while the work of formation approaches zero at the spinodal~\cite{Iwamatsu08}.  On the other hand, both the size and the work of formation of the critical cavity approaches zero at the spinodal.  Even though the work of formation of critical cavity is higher than that of the critical bubble as shown in Figs.~\ref{fig:4}, the sub-critical cavities smaller than the critical cavity could be easier to form than the much larger and diffuse critical bubble near the spinodal.  Then the bubble formation might be preceded by the (sub-critical) cavity formation at least near the spinodal, and the homogeneous nucleation of bubble might be in fact the heterogeneous nucleation where the homogeneous bubble nucleation starts from the cavity formation.  As the size of the sub-critical cavity grows, it transforms into the bubble by accommodating the liquid molecule into the cavity and lowers its free energy (work of formation) before reaching the critical cavity.  In such a case the phase separation of stretched liquid to vapor occurs via the gradual heterogeneous bubble nucleation around the sub-critical cavity rather than the explosive critical cavity formation that corresponds to the stability limit of metastable liquid.  In fact, recent molecular dynamics~\cite{Zahn04} and Monte Carlo~\cite{Neimark05} studies of bubble formation and boiling show that the bubble nucleation is initiated by the spontaneous formation of vacuum cavity.

In the previous paper, we have pointed out that this triple-parabolic free energy has an artifact that the isothermal compressibility remains constant as the parameter $\lambda_{i}$ is fixed in Eq.~(\ref{eq:2}), which reflects in the finite and constant size of the critical bubble represented by the constant dividing radius $r_{D}$ or the matching radius $r_{B}$.  The same behavior is expected when we use the double-parabola model~\cite{Iwamatsu93}. It is expected, in fact, the isothermal compressibility $\kappa$ of the liquid phase diverges as~\cite{Kalikmanov04}
\begin{equation}
\kappa \propto \lambda_{2}^{-1} \propto \left(1-\frac{\Delta\mu}{\Delta\mu_{\rm spin}}\right)^{-1/2}.
\label{eq:xx}
\end{equation}
from Eq.~(\ref{eq:8}) of the quartic $\phi^{4}$ field model~\cite{Unger84}.  Then the size of the critical bubble should diverge as~\cite{Cahn59,Klein83,Kalikmanov04}
\begin{equation}
r \propto \left(1-\frac{\Delta\mu}{\Delta\mu_{\rm spin}}\right)^{-1/4}
\rightarrow \infty
\label{eq:xx2}
\end{equation}
when $\Delta\mu/\Delta\mu_{\rm spin}\rightarrow 1$ as the density profile of the critical bubble is given by Eq.~(\ref{eq:16}) with $R=0$ and $r_{B}$ is determined from $\lambda_{2}$~\cite{Iwamatsu08} in the spinodal regime.  Then, the story for the critical bubble in real materials would be slightly different from that of our triple-parabolic model, in particular, in the spinodal regime.  On the other hand, since the density profile of the critical cavity is always given by Eqs. ~(\ref{eq:13}) and (\ref{eq:14}) of the CNT regime, the size of the critical cavity $R$ will be relatively insensitive to the $\lambda_{2}$.

It is clear from Fig.~\ref{fig:2} and Eq.~(\ref{eq:xx2}) that there exists a minimum size of the critical bubble, while the size of the critical cavity approaches zero.  Furthermore, the critical size $R$ of the cavity is smaller than the size of the critical bubble represented by the dividing radius $r_{B}$ from Fig.~\ref{fig:6}(a).  Even though the critical size of the cavity is smaller than the bubble, it is also clear from Fig.~\ref{fig:6}(b) that the work of formation of the critical cavity is always larger than that of the critical bubble.

Since both the size of the critical cavity $R$ and the work of formation $W^{*}$ approaches zero as we move toward the spinodal, it is expected that the sub-critical cavity rather than much larger critical bubble could easily be formed and induce phase transformation from metastable liquid to vapor near the spinodal. The previous picture of homogeneous nucleation assumes that the long-wavelength fluctuation induces the critical bubble formation.  Near the spinodal, such a homogeneous nucleation called spinodal nucleation would occur cooperatively and the sudden and the rapid phase transformation could occur as the system becomes nearly unstable.  Our simple calculation using DFT model indicates that this scenario of phase transition near the (mean-field) spinodal may not be true and the sub-critical cavity rather than the critical bubble could play some role to induce the rapid spinodal nucleation.

\section{Conclusion}
\label{sec:sec4}

In this work, we have studied the size and the work of formation of the critical cavity using a generic square-gradient density-functional model with triple-parabolic free energy.  We have used this model as it is generic in the sense that it does not depend on the interatomic potential and it has already been used to study the scaling properties of the critical bubble~\cite{Iwamatsu08}.  We pay particular attention to the comparison of the critical cavity with the critical bubble of the homogeneous nucleation to check the  conjecture made by Corti and coworker~\cite{Punnathanam03} for the Lennard-Jones fluid. 

Aside from the fact that our DFT model assumes constant compressibilities $\lambda_{i}, i=0, 1, 2$ in Eq.~(\ref{eq:3}) and a constant square-gradient coefficient $c$ in Eq.~(\ref{eq:2}), it seems certain from our numerical results that the two conjecture made by Punnathanam and Corti~\cite{Punnathanam03} on the size and the work of formation of the critical cavity are valid.  From the behavior of the critical cavity near the spinodal, sub-critical small cavity seems easier to form than the critical bubble in stretched fluid with negative pressure near the spinodal.  This conclusion for the sub-critical cavity and critical bubble near the spinodal remain correct even if we include the diverging isothermal compressibility at the spinodal predicted from the $\phi^{4}$-field theory\cite{Debenedetti96,Unger84,Kalikmanov04}. Then it could be argued that the sub-critical cavity~\cite{Punnathanam03,Uline07} plays some role to induce critical bubble nucleation in the stretched fluid at least near the spinodal.  If the sub-critical cavity plays the crucial role in liquid-to-vapor phase transformation (vaporization), the real picture of the phase transformation should be more complex.  Then the real picture of phase transformation is totally different from the traditional picture based on the homogeneous nucleation where the phase transformation occurs through the formation of the critical bubble from the fluctuation and subsequent growth of the bubble.  It would be, however, difficult experimentally to judge whether the phase transformation near the spinodal proceeds via the cavity or bubble as the liquid phase becomes nearly unstable near the spinodal where the fluctuation plays the dominant role and smears the detail of the phase transformation. 

Finally, the dynamics of cavitations would be more interesting for practical purposes. Various numerical methods such as the Monte Carlo~\cite{Leung03}, the molecular dynamics~\cite{Okumura03} and the lattice Boltzmann method~\cite{Sukop05} have been developed to study the dynamic of cavity or bubble formation and the evaporation.  Recently, we have developed a numerical method based on the time dependent Ginzburg-Landau model combined with the cell dynamics method to study the dynamics of nucleation in various situations~\cite{Iwamatsu07,Iwamatsu08a}.  It will be interesting to use this cell dynamics method to study the dynamics of cavity or bubble formation.

\end{document}